\documentclass[11pt]{article}
\usepackage[utf8]{inputenc}
\usepackage[T1]{fontenc}
\usepackage{lmodern}
\usepackage{geometry}
\usepackage{microtype}
\usepackage{amsmath,amssymb}
\usepackage{xcolor}
\usepackage{xurl}
\usepackage{graphicx}
\usepackage{hyperref}
\usepackage{enumitem}
\hypersetup{colorlinks=true, linkcolor=blue, citecolor=blue, urlcolor=blue}
\title{Hound: Relation-First Knowledge Graphs for Complex-System Reasoning in Security Audits}
\author{Bernhard Mueller}
\date{September 2025}

\newcommand{\keywords}[1]{\par\noindent\textbf{Keywords:} #1}

\begin{document}
\maketitle
\begin{abstract}
Hound introduces a relation-first graph engine that improves system-level reasoning across interrelated components in complex codebases. The agent designs flexible, analyst-defined views with compact annotations (e.g., monetary/value flows, authentication/authorization roles, call graphs, protocol invariants) and uses them to anchor exact retrieval: for any question, it loads precisely the code that matters (often across components) so it can zoom out to system structure and zoom in to the decisive lines. A second contribution is a persistent belief system: long-lived vulnerability hypotheses whose confidence is updated as evidence accrues. The agent employs coverage-versus-intuition planning and a QA finalizer to confirm or reject hypotheses. On a five-project subset of ScaBench\cite{scabench}, Hound improves recall and F1 over a baseline LLM analyzer (micro recall 31.2\% vs. 8.3\%; F1 14.2\% vs. 9.8\%) with a modest precision trade-off. We attribute these gains to flexible, relation-first graphs that extend model understanding beyond call/dataflow to abstract aspects, plus the hypothesis-centric loop; code and artifacts are released to support reproduction.
\end{abstract}

\section*{Contributions}
\begin{itemize}[leftmargin=*]
  \item \textbf{Flexible relation-first graphs that amplify LLM code comprehension:} analyst-defined, multi-scale graphs that capture system architecture and key aspects (authentication roles, monetary/value flows, call graphs, invariants), \emph{iteratively refined with typed annotations} (observations, assumptions) to enable exact cross-component retrieval and reasoning.
  \item \textbf{Belief system:} long-lived vulnerability hypotheses with confidence updated as evidence accrues, providing continuity across sessions and guiding planning/review.
\end{itemize}

% (Outline removed; Contributions below provide orientation.)
\section{Challenges}
LLM-based code analysis faces several significant challenges that Hound targets directly.

\paragraph{Scalability and long-range context.}
Auditing large, complex repositories requires multi-file navigation, cross-cutting state tracking, and sustained intent over many steps. Current models struggle with repository‑scale context and long‑horizon reasoning \cite{jimenez2023swebench}. Limited context windows fragment understanding; important cross‑file interactions can be missed unless systems gather distributed evidence and reason across files. Empirical work shows that broader, structured context improves outcomes: agents that explore many more files per query (“deep research”) achieve markedly better results on large systems \cite{singh2025coderesearcher}. Graph‑ and flow‑aware methods likewise improve multi‑hop code tasks over naive similarity search \cite{ouyang2024repograph,wang2025coderagbench,liu2025codexgraph}.

Much recent research has focused on semantic views, such as knowledge graphs, code graphs, and AST‑guided indices, to help LLMs maintain a consistent picture of a codebase and focus on relevant regions. Graph‑guided agents improve fuzzing and bug discovery by steering exploration \cite{xu2024codegraphgpt}, and AST‑guided search improves localization and fixes in large projects \cite{zhang2024autocoderover}. Repository graph databases enable precise, query‑driven retrieval (e.g., Cypher over code relationships), significantly boosting repository‑level performance \cite{liu2025codexgraph}. Industry reports echo these trends, highlighting knowledge‑graph patterns for AI systems and codebases.

\paragraph{Reliability and truthfulness.}
Unconstrained generation can yield hallucinated code facts or false positives when evidence is thin or missing. In auditing, this undermines trust: models may assert issues that do not exist or miss subtle vulnerabilities that do. Studies report concrete failure modes (e.g., “package hallucinations”) and recommend grounding claims in verifiable code evidence \cite{spracklen2025pkg}. Ensuring that explanations are supported by precise references remains a core requirement for credible AI-assisted audits.

Combining LLMs with program analysis improves precision at scale. Repo‑level auditors pair step‑wise navigation with data‑/control‑flow checks to validate suspected bugs and reduce false positives \cite{guo2025repoaudit}. Surveys and position papers argue that advancing the state of the art requires scaling context, integrating external knowledge, and tightening correctness guarantees \cite{jelodar2025survey}. Recent work also shows a reciprocal link between code understanding and reasoning skills, motivating approaches that strengthen both \cite{yang2025codetothink}.

\medskip
\section{Hound's Approach}
Hound uses a graph-based, agentic architecture that enables multi-scale (zoom-in/zoom-out) understanding of large systems, persistent memory, precise retrieval, and structured reasoning. Crucially, its graphs are \emph{flexible}: beyond call/dataflow, they model the aspects auditors actually reason about (authentication/authorization roles, monetary/value flows, call graphs, invariants, protocol phases), expanding what the model can comprehend and test.
\subsection{Addressing Inadequate Memory and Iterative Reasoning}
Expert auditors do not keep a single, frozen view of a system in mind. They sketch a mental map, revise it when new interactions come into view, and carry that understanding forward across days or weeks. Hound captures this habit with \emph{living knowledge graphs} that persist across sessions. Each audit begins with a high-level SystemArchitecture graph and then expands into focused ``aspect'' graphs (authorization, asset flow, state mutation, upgrade lifecycle, and more). Nodes and edges carry typed annotations (observations, assumptions) with compact evidence links back to the precise code slices that justified them, so the graph becomes a durable, self-explaining record rather than a transient prompt.

This persistent representation is complemented by light-weight, project-wide stores that reflect how humans track progress and open questions. A coverage index remembers which nodes and code cards have been touched; a plan ledger records recurring investigation frames; a hypothesis store preserves the evolving belief set with provenance. Within a single investigation loop, Hound also manages working memory: once context approaches its budget, older interaction history is compressed into concise ``memory notes,'' while the most recent actions are kept verbatim. The net effect is an agent that resumes quickly, remembers what mattered, and steadily refines its internal model instead of starting from scratch.
\subsection{Addressing Low-Precision Retrieval}
When an auditor studies a function or a specific aspect such as a monetary flow, they first identify the functions and paths that carry that aspect and then read the exact lines that implement or use them, not an approximate snippet that happens to be nearby. Hound mirrors this attention to precision. The system ingests repositories into byte-accurate \emph{cards} (file-relative slices with character offsets). Graph nodes and edges may reference these card IDs as evidence, and graphs are iteratively refined with typed annotations (observations, assumptions) that capture cross-component invariants and intent. When the agent asks to ``load'' a node, Hound resolves the cards referenced by that node and its incident edges (where present), orders them deterministically by file and offset, and presents the minimal code necessary for the reasoning step.

Because retrieval is anchored in graph evidence rather than similarity search, context stays tight and reproducible. In contrast to vector-embedding RAG (e.g., cosine over code or descriptions), which often surfaces lookalike interfaces and loosely related snippets, Hound follows typed edges in task-specific graphs to determine \emph{exactly} what else is relevant (across components when needed) and loads only those precise slices. This keeps prompts clean, reduces distraction from off-target context, and makes the chain of reasoning auditable end to end.
\subsection{Addressing Absence of Structured Belief Refinement}
Good auditors state what they think is wrong, where to look, and how confident they are, then they change their mind when the evidence demands it. Hound turns that discipline into first-class data. Findings begin as \emph{hypotheses} with explicit fields (title, type, severity, confidence, linked node IDs, and reasoning). As evidence accrues, confidence is adjusted and status moves through proposed, investigating, supported, refuted, and finally confirmed or rejected. Crucially, hypotheses are long-lived across sessions: they persist, accumulate evidence, guide planning, and provide continuity for teams. All updates are atomic and auditable.

To seed high-quality leads, a senior ``Strategist'' model reviews compact, code-grounded context and emits small batches of targeted investigations or candidate hypotheses. This step can run in a single pass or a conservative two-pass mode that self-critiques the first list to reduce false positives. Later, a separate ``Finalizer'' model performs QA over full source context and records a verdict with justification. Together, these roles reproduce the cadence of a seasoned team: a senior auditor proposes promising lines of inquiry, a junior digs into the code, and a reviewer signs off when the claim is sufficiently evidenced.
\subsection{Addressing Inability to Reason Across Multiple Abstraction Layers}
Many impactful bugs hide in the seam between architectural intent and concrete implementation. Human auditors move fluidly between the two: they sketch flows and invariants at the system level, then ``zoom'' into just the right functions and state transitions to test whether reality matches the sketch. Hound operationalizes that zoom. Aspect graphs are intentionally diverse (system overview first, then domain-relevant lenses like authorization maps, asset movement, or reentrancy surfaces) so the agent can reason over the right abstraction at the right time.

When a lead emerges, the agent pivots from the abstract graph to exact code by loading only the nodes (and thus cards) that bear on the question. Observations and assumptions live next to those nodes and edges, so contradictions surface naturally: a stated invariant that is not enforced where expected becomes a saliency signal for deeper analysis. Planning adapts with coverage: early passes favor broad, component-level sweep; later ones concentrate on high-impact suspicions guided by value at risk and mismatched assumptions. In effect, Hound gives the model the same handholds a human uses to climb the abstraction ladder: clear maps, precise footholds, and a disciplined way to mark what was learned and what still needs to be true.
\subsection{Addressing Brittle Dependence on Static Analysis Tooling}
\textit{Intuition:} Top-performing auditors in competitive contests rely primarily on an IDE and careful reading. Heavy static tools help occasionally, but human judgment, mental models, and targeted slicing uncover most impactful bugs.

Traditional pipelines lean on language-specific parsers and heavyweight static analyzers. They excel when the grammar is familiar and the build succeeds, but they can become brittle across heterogeneous stacks or unusual project layouts. Hound takes a deliberately different path. It builds its models from \emph{raw code cards} and agent-discovered graph schemas, without depending on ASTs or toolchain hooks. Nodes and edges are typed by intent (e.g., \textit{function}, \textit{role}, \textit{storage}, \textit{calls}, \textit{guarded\_by}) rather than by parser output, and every structure is anchored to evidence slices by character offset. This makes the representation resilient across languages and frameworks while remaining precise enough to point to the exact lines when it is time to verify.

The language-agnostic stance decouples from static tooling. Static analyzers can plug in as optional evidence sources, but the core loop (graph discovery, hypothesis formation, and belief revision) remains robust even when compilers do not cooperate or when the codebase mixes languages. In practice, this lets Hound traverse unfamiliar terrain the way a human would: sketching relationships first, then tying them back to concrete text where needed.

\subsection{Addressing Lack of Effective Audit Planning Heuristics}
Experienced reviewers begin with a rapid breadth-first scan to map components, interactions, and value flows, forming and testing hypotheses as they go, rather than reading every line end to end. Hound separates \emph{planning} from \emph{verification} to support this cadence. The Strategist is shown a \emph{rich graph-only view}: multiple aspect graphs with compact annotations (observations and assumptions), but \emph{without} code blobs. In that sparse, high-signal view, contradictions and value at risk patterns stand out: an invariant noted in one place but unenforced along a critical edge; a permission check attached to most, but not all, routes into a value-moving function. From this, the Strategist proposes a small set of investigations with ``why now'' and clear exit criteria, switching between broad coverage and intuition-driven deep dives as project coverage improves.

\paragraph{Two-phase planning in practice.} Hound runs planning in two phases: \textsc{Coverage} (\emph{sweep}) until overall coverage reaches a threshold (default \(p^*\!\approx\!90\%\)), then \textsc{Intuition} (\emph{saliency}) for deep, high-impact leads. The Strategist's \texttt{plan\_next} uses phase-specific prompts; in auto mode it selects the phase from the coverage summary, while runners may override the phase explicitly. This mirrors a human audit: first sweep to map components and pick low-hanging fruit; then pursue cross-component, high-value suspicions.

Only then does the Scout retrieve code, precisely and minimally, by loading the evidence cards for the targeted nodes and their incident edges. Hypotheses formed in this execution phase inherit the plan’s focus but are grounded in exact slices, with confidence adjusted as evidence accumulates. The cycle is recorded in a coverage index and plan ledger to avoid wasteful repetition while still allowing deliberate revisits when new evidence appears. This rhythm mirrors a well-run human audit: think with the maps; verify with the text.
\section{Architecture and Data Model}
Hound comprises: (i) an agent-designed multi-graph model of the codebase, (ii) a junior \emph{Scout} that loads code and prepares context, (iii) a senior \emph{Strategist} that plans investigations and proposes hypotheses, and (iv) a \emph{Finalizer} for QA.

Data artifacts (per project):
\begin{itemize}[leftmargin=*]
  \item Manifest and cards: repository ingested into byte-sliced \emph{cards} with file-relative offsets; a manifest tracks files and offsets to reconstruct spans when needed.
  \item Graphs: a collection of named graphs with nodes, edges, and optional references to card IDs (\texttt{source\_refs} on nodes, \texttt{evidence} on edges).
  \item Referenced-card store: retains only cards actually referenced by any graph for reproducible retrieval.
  \item Hypotheses: a persistent store of hypotheses with confidence and status maintained per hypothesis (not per graph node).
  \item Coverage and plans: per-project coverage index and per-session plan store, plus a project-wide ledger of normalized plan frames.
\end{itemize}

Retrieval is reference-driven: when the Scout loads a node, it collects the node's `source\_refs` plus `evidence` from incident edges, orders the corresponding cards deterministically by `(relpath, char\_start)`, and presents only those slices. By following graph relations, the agent determines and loads exactly the code needed (often across components) for the current question; loaded nodes return precise, question-relevant snippets rather than broad file dumps.

\subsection{Formal Model}
We formalize the core objects the system manipulates: byte-sliced cards, agent-designed graphs with optional references to those cards, hypotheses as belief records with confidence/status, and lightweight planning/coverage state.
Let a repository be byte text \(R\) partitioned into cards \(C=\{c_i\}\), where each \(c_i=(\mathrm{id},\,\mathrm{relpath},\,[\mathrm{cs},\mathrm{ce}))\) with content \(s(c_i)=R[\mathrm{cs}:\mathrm{ce}]\). Cards for a file are non-overlapping slices chosen for coverage and locality.

For a graph \(g\), define
\[
g=(V, E, \tau_V, \tau_E, \rho_V, \rho_E, O, A),
\]
where \(\tau_V:V\to\mathcal{T}_V\) and \(\tau_E:E\to\mathcal{T}_E\) are type maps (e.g., function, storage, role; calls, guarded\_by, writes). Optional reference maps \(\rho_V:V\to 2^C\) and \(\rho_E:E\to 2^C\) attach card IDs to nodes and edges, respectively. Each node \(v\in V\) may carry observations \(O_v\subseteq\mathcal{D}\) and assumptions \(A_v\subseteq\mathcal{D}\) as sets of textual statements; confidences are \emph{not} tracked on nodes (they belong to hypotheses).

\paragraph{Intuition-mode planner (formal).} Given graphs \(G\), coverage \(\Gamma\), hypothesis map \(H\), and plan ledger \(L\), the intuition planner returns up to \(n\) investigations
\[
\begin{aligned}
\Pi_{\mathrm{int}}(G,\Gamma,H,L,n) &\;\to\; \{\, i_k \,\}_{k\le n},\\
i &\;=\; (\text{goal},\,\text{category},\,\text{focus\_areas},\\
&\quad \text{priority}\in\{1..10\},\,\text{expected\_impact}\in\{\text{high,med,low}\},\\
&\quad \text{reasoning}).
\end{aligned}
\]
Items must not repeat frames already present in the per‑session PlanStore \(\mathcal{F}\) or project ledger \(L\). Selection follows a phase‑specific preference order (implemented via LLM prompts, not numeric scoring):
\begin{itemize}[leftmargin=*]
  \item maximize saliency (contradictions between assumptions and observations),
  \item prefer higher expected impact (value at risk),
  \item prefer novelty w.r.t. coverage (unvisited nodes/cards) and existing hypotheses/plans,
  \item maintain diversity of focus areas to avoid tunneling.
\end{itemize}
This is realized in code by \texttt{Strategist.plan\_next(..., phase='Saliency')} which emits structured items with \texttt{priority} and \texttt{expected\_impact}; the agent enforces no‑repeat constraints using \(\mathcal{F}\) and \(L\).

Coverage state is \(\Gamma=(\chi_V,\chi_C)\), with \(\chi_V:V\to\mathbb{N}\) and \(\chi_C:C\to\mathbb{N}\) recording visits (node/card touches). A plan frame is a tuple \(f=(\text{goal},\,\text{category},\,\text{focus\_areas},\,\text{status})\); the per-session PlanStore is a finite set \(\mathcal{F}\) of frames. A project-wide ledger maps normalized frames to counts and provenance.

Let \(p(\Gamma)\in[0,1]\) denote coverage (e.g., percent of nodes or cards visited). Planning selects a phase \(m\in\{\textsc{Coverage},\textsc{Intuition}\}\) by
\[
m\;=\;\begin{cases}
\textsc{Coverage} & \text{if } p(\Gamma) < p^*\\
\textsc{Intuition} & \text{otherwise}\,,
\end{cases}\qquad p^*\approx 0.9\ \text{(default, configurable).}
\]
The planner \(\Pi(G,\Gamma,H,L,m,n)\) returns up to \(n\) investigations. In the implementation, \(\Pi\) is realized by an LLM with phase-specific prompts rather than an explicit numeric objective. It follows a phase-dependent preference order \(\prec_m\):
\begin{itemize}[leftmargin=*]
  \item \textsc{Coverage}: prefer previously unvisited, medium‑granularity components; prioritize broadly useful or high‑impact areas for the project; avoid repeats already in \(\mathcal{F}\) or \(L\).
  \item \textsc{Intuition}: prefer high expected impact (value at risk), salient contradictions between assumptions and observations, cross-component interactions, and novelty w.r.t. existing hypotheses and plans.
\end{itemize}
Returned items include fields like \texttt{priority} and \texttt{expected\_impact} that the Strategist provides; no weighted sum is computed in code.

Hypotheses form a finite map \(H\) from identifiers to tuples \(h=(\text{title},\text{type}, S\subseteq V, \text{severity}, q\in[0,1], \text{status}, E\subseteq C, \text{reason})\). A senior operator \(\Delta\) proposes candidates from graph‑only context; verification retrieves the referenced cards for \(S\) to collect evidence \(e\subseteq C\). Updates include: \(\mathrm{propose}:H\to H\), \(\mathrm{add\_evidence}:H\times C\to H\), and \(\mathrm{adjust\_q}:H\times[0,1]\to H\) which may set \(\text{status}\in\{\text{proposed},\text{investigating},\text{supported},\text{refuted},\text{confirmed},\text{rejected}\}.\)

When prompt tokens exceed a threshold \(\tau B\), a summarizer \(\Sigma\) compacts stale history into memory notes while preserving the last \(\kappa\) actions verbatim.

\subsection{Architecture and Algorithms}
\paragraph{Cards and Indexing.} Repositories are sliced into byte-accurate cards with \(\langle\mathrm{relpath},\mathrm{cs},\mathrm{ce}\rangle\) metadata. An index combines card metadata with the ingestion manifest to allow exact span reconstruction when content is omitted.

\paragraph{Graph Builder.} The builder discovers a \emph{SystemArchitecture} graph and diverse aspect graphs via an LLM prompt, then iteratively refines each graph with strictly typed updates (new nodes/edges and compact node updates). Nodes and edges may include references to cards; referenced cards are persisted in a separate card store. Discovery and refinement are token-aware (context estimates, sampling) and stop early when no new nodes/edges are admitted.

\medskip\noindent\textit{Discovery.} Given the manifest and a sample of code cards sized to the active model's context, the builder requests exactly \(k\) graphs (\(k\ge1\)); the first is forced to \emph{SystemArchitecture}. The response schema is:
\begin{itemize}[leftmargin=*]
  \item GraphDiscovery: \texttt{\{ graphs\_needed: [GraphSpec], suggested\_node\_types: [str], suggested\_edge\_types: [str] \}}
  \item GraphSpec: \texttt{\{ name: str, focus: str \}}
\end{itemize}
If graphs are forced via CLI (single-graph spec), discovery is skipped and that graph is created immediately.

\medskip\noindent\textit{Iterative Build.} For each iteration, the builder asks for an update to one target graph with the schema:
\begin{itemize}[leftmargin=*]
  \item GraphUpdate: \texttt{\{ target\_graph: str, new\_nodes: [NodeSpec], new\_edges: [EdgeSpec], node\_updates: [NodeUpdate] \}}
  \item NodeSpec: \texttt{\{ id: str, type: str, label: str, refs: [card\_id] \}}
  \item EdgeSpec: \texttt{\{ type: str, src: str, dst: str, refs: [card\_id] \}}
  \item NodeUpdate: \texttt{\{ id, description?, properties?, new\_observations?, new\_assumptions? \}}
\end{itemize}
New nodes are added once (merging \texttt{refs}); edges are de-duplicated by \(\langle\text{src},\text{dst},\text{type}\rangle\) while merging \texttt{refs}. Updates merge descriptions/properties and append observations/assumptions. A \emph{refine-only} mode restricts new nodes to those attached to existing nodes through proposed edges and caps their number conservatively.

\medskip\noindent\textit{Token-aware sampling and early stop.} Discovery samples cards to stay within the model's context; the builder logs token counts and warns when close to the limit. During refinement, the builder stops early when consecutive iterations add no nodes or edges (after a minimal warm-up), and saves graphs incrementally after each applied update.

\medskip\noindent\textit{Persistence and provenance.} Graphs persist with stats and metadata; only cards referenced by any node or edge are retained in a compact store for reproducible retrieval. A summary records totals across graphs. The builder also surfaces disconnected (orphan) nodes per graph to guide refinement.

\paragraph{Agent Loop.} \textit{Intuition:} A junior auditor explores deliberately, gathering minimal, relevant code to brief a senior and postponing verdicts until evidence is assembled. 
\textit{Implementation:} The Scout executes a strict action loop validated by schemas: \texttt{load\_graph}, \texttt{load\_nodes}, \texttt{update\_node}, \texttt{form\_hypothesis}, \texttt{update\_hypothesis}, and \texttt{complete}. Node loading requires an explicit graph name and exact node IDs; the agent collects code from node \texttt{source\_refs} and incident-edge \texttt{evidence}, ordered deterministically, yielding exact snippets scoped to the current question. \texttt{update\_node} adds human-readable observations/assumptions to nodes. Context building mirrors an auditor's notebook: investigation goal, steering notes, available graphs, compressed memory notes, full SystemArchitecture (compact), optionally loaded graphs, cached node IDs, recent actions, and existing hypotheses (grouped by type with status and confidence). When context approaches a budget, older history is compressed into memory notes while keeping the last \(\kappa\) actions verbatim; token usage is reported inline.

\paragraph{Strategist and Planning.} \textit{Intuition:} A senior auditor first sweeps broadly to understand the system, then pivots to deep, high-impact hunches guided by contradictions and value at risk. \textit{Implementation:} The Strategist consumes a graph-only context and plans with two modes: \emph{Coverage} (systematic component sweep) and \emph{Intuition/Saliency} (deep dives on high-risk suspicions). The planning API \texttt{plan\_next} returns prioritized \emph{Investigations} with \emph{why now} and \emph{exit criteria}. Investigations are structured items with fields \{goal, category, focus\_areas, priority, expected\_impact, reasoning\} that direct attention to the most promising aspects next. For deep analysis, \texttt{deep\_think} emits structured hypothesis candidates (title, type, severity, confidence, node\_ids, reasoning). Optional two-pass review self-critiques an initial batch; LLM-assisted deduplication filters near-duplicates against existing hypotheses. Runners persist plans in a per-session PlanStore and normalize frames in a project-wide PlanLedger.

\paragraph{Reference-Driven Retrieval.} Executions retrieve code by node by following explicit node/edge references to gather only the cards relevant to the current question, ordered by file and offset. This produces exact, minimal slices (often spanning multiple components) directly from the graph. In contrast, similarity-based retrieval (e.g., cosine over code or descriptions) tends to surface lookalike interfaces or loosely related snippets, diluting analysis with off-target context.

\paragraph{Belief Lifecycle and QA.} \textit{Intuition:} Human auditors float leads with a confidence slider, attach exhibits to the case file, and ask a reviewer to render a verdict over the full record.
\textit{Implementation:} Hypotheses are first-class, long-lived objects with fields (title, type, severity, confidence \(q\), status, node\_refs, evidence, reasoning, properties). Confidence/status are only tracked at the hypothesis level (not on graph nodes). Store operations include: \texttt{propose} (duplicate-safe by title), \texttt{add\_evidence} (appends evidence and heuristically updates status: investigating/supported/refuted), and \texttt{adjust\_confidence} (updates \(q\) and marks \texttt{rejected} when \(q\le 0.1\)). The Scout forms hypotheses from Strategist output with deduplication and populates \texttt{properties} (graph name, source files, affected functions) from referenced nodes and heuristic path extraction. Finalization loads relevant repository files (up to a small cap) and prompts a QA model to issue a JSON verdict \{\texttt{confirmed}\,\texttt{rejected}\,\texttt{uncertain}\} with reasoning; confirmed items are set to \texttt{status=confirmed}, \(q=1.0\); rejected items are set to \texttt{status=rejected}, \(q=0.0\); uncertain hypotheses remain pending with an explanatory note.

\paragraph{Coverage and Planning Stores.} \textit{Intuition:} An expert team keeps two boards: a coverage board (what's been examined) and a plan board (what to do next), while a project log tracks recurring frames across sessions.
\textit{Implementation:} The CoverageIndex records visited node/card IDs with counts and timestamps; it reports coverage at node and card granularity and can seed from prior session activity. Plans are persisted per session in a PlanStore as frames with status \{\texttt{planned,in\_progress,done,dropped,superseeded}\} and history; a project-wide PlanLedger normalizes frames (question + artifact refs), aggregates counts, and records which sessions/models have proposed them. All stores inherit atomic, process-safe updates from a common file store with OS locks and per-file thread locks, enabling multi-session collaboration without corruption.

\paragraph{Context and Budgets.} The agent reports token usage per call and overall context utilization; compression triggers at a configurable fraction of the maximum context and keeps only the most recent actions verbatim (default threshold \(\tau\approx0.75\), recent actions \(\kappa\approx5\)). Model selection is profile-based and provider-agnostic (OpenAI, Anthropic, Gemini, XAI, DeepSeek, Mock).

\subsection{Adaptive Planning: Coverage $\to$ Intuition}
Let total node coverage be
\[
 p\;=\;w_V\,\frac{|V_{\text{visited}}|}{|V|}\; +\; w_C\,\frac{|C_{\text{visited}}|}{|C|},\quad w_V,w_C\ge 0,\;w_V+w_C=1,
\]
computed from \(\Gamma=(\chi_V,\chi_C)\). Define a target threshold \(p^*\) (e.g., \(0.9\)). The planning policy interpolates between a \emph{coverage objective} and an \emph{intuition objective} via a mixing coefficient
\[
 \lambda\;=\;\min\{1,\max\{0,\,\frac{p-p_0}{p^*-p_0}\}\},\quad 0\le p_0<p^*\,.
\]
Investigations \(i\) are produced by the planner according to the active phase (\textsc{Coverage} or \textsc{Intuition}). In practice, selection is LLM‑driven with guardrails rather than an explicit numeric objective: avoid repeats in \(\mathcal{F}\)/\(L\); maintain breadth in \textsc{Coverage}; and in \textsc{Intuition}, prefer salient contradictions, higher expected impact, and novelty w.r.t. coverage and existing hypotheses.

\paragraph{Exit criteria.} Each investigation carries explicit criteria: terminate when (i) a decisive hypothesis surpasses confidence \(q^*\), (ii) contradiction is resolved (evidence supports or refutes), or (iii) retrieval/compute budget for \(i\) is exhausted. Outcomes update \(\Gamma\) and the hypothesis store, and future \(\Pi\) calls adapt via the new \(p\) and ledger entries.

\subsection{Audit Strategy in Practice}
\paragraph{Low-hanging fruit via Sweep.} Early in an engagement, the goal is to gain meaningful coverage and surface obvious issues quickly. Hound groups nodes into medium-sized components (contracts, modules) and selects investigations that maximize new node/card visitation per token. Exclusions (tests, mocks, vendored code) and trust-boundary heuristics keep focus on production paths. The resulting items load only the evidence cards for the selected component’s frontier nodes, allowing rapid iteration and fast elimination of shallow bugs.

\paragraph{Intuition-guided deep dives.} Once basic structure is internalized (\(p\ge p^*\)), the Strategist pivots to a graph-only context: aspect graphs and annotations, plus a compact hypotheses summary with statuses. From this high-signal view, it follows up on interesting leads (e.g., a value-moving function guarded on most but not all paths) and targets contradictions between assumptions and observations. Plans specify ``why now'' and exit criteria; execution retrieves only the cards needed to test the claim, adjusts confidences, and records coverage.

\paragraph{Intuition mode exploration.} The Strategist starts an investigation with a clear goal (high‑level or detailed). The Scout then explores the graphs to select the specific nodes and paths most relevant to that goal and loads their referenced code (node \texttt{source\_refs} plus incident edge \texttt{evidence}). When additional depth or hypotheses are needed, the Scout escalates back to the Strategist for guidance and hypothesis generation. Examples: early on, direct attention to high‑impact aspects such as monetary/value flows; later, follow up on a prior hypothesis or zoom in on a specific algorithm; or, upon spotting a mismatch between an invariant and observations, “go down the rabbit hole” to resolve the contradiction.

\paragraph{Economical model roles.} Planning and hypothesis generation demand the strongest reasoning model; execution and code loading can run on a smaller model. Hound separates these concerns: a \textit{Strategist} profile (with optional \texttt{plan\_reasoning\_effort} and \texttt{hypothesize\_reasoning\_effort}) produces compact, high-value plans and candidates; a \textit{Scout} profile executes actions, loads nodes, and proposes/updates hypotheses with strict, typed parameters. This split yields senior-level guidance at junior-level cost, while preserving determinism through exact node IDs and reference-driven retrieval.

\paragraph{Follow-ups on findings.} The Strategist sees current hypotheses with statuses and confidences, and can prioritize confirm/refute investigations for high-impact but uncertain items. Optional two-pass review prefers fewer, stronger candidates. Finalization then reviews high-confidence items with direct access to repository files (reconstructing spans when needed), while leaving genuinely undecidable items open with intermediate confidence.

\subsection{Concurrent Teams and Collaboration}
Hound’s storage layer enables concurrent audits by multiple agents and model profiles against the same project. Hypotheses carry \texttt{created\_by} and \texttt{session\_id} tags and are written via atomic, lock-guarded updates; the coverage index and plan ledger aggregate activity across sessions. This allows parallel teams to share a single belief set while pursuing different angles. Lightweight ``steering'' is supported through a project inbox: external notes can preempt a running investigation and request a replan, mirroring how human leads redirect effort during a live engagement.

Because retrieval is reference-driven (where references exist), concurrent work cannot trample context: actions resolve to cards through \(\rho\) when available, and per-process locks ensure store integrity. This design admits diverse orchestration patterns, from one strong Strategist guiding several Scouts to multiple independent cells, without introducing a central database or coordination service.

\subsection{Standard Audit Phases}
\textbf{1) Graph Build.} The audit begins by discovering \emph{SystemArchitecture} and complementary aspect graphs and persisting them with their evidence map. Card indices and manifests are written alongside to support exact reconstruction of referenced spans.

\textbf{2) Coverage Sweep.} Planning prioritizes medium-sized components across the codebase, maximizing new node/card visitation under budget. Investigations are expressed with ``why now'' and exit criteria and recorded into the plan ledger and coverage index.

\textbf{3) Intuition Deep Dives.} As coverage improves, planning shifts toward high-impact suspicions guided by contradictions and value-at-risk. This phase can run indefinitely: cycles of (plan \(\to\) retrieve \(\to\) hypothesize \(\to\) update) continue until time or budget limits are reached, or until gains saturate.

\textbf{4) Finalization (QA).} High-confidence hypotheses are reviewed over full source context. The finalizer reads the project manifest to access repository files directly, reconstructing exact spans when necessary to ensure nothing material is out of view. Verdicts and reasoning are written back to the hypothesis store.

\textbf{5) Reporting.} A professional report is generated from confirmed findings, with optional inclusion of open items and attached proof-of-concept artifacts when available.

\subsection{Conservative QA and Open Findings}
Hound allows hypotheses to remain undecided when the available evidence is insufficient or when necessary code is unavailable. In such cases, items retain an intermediate confidence (e.g., \(q\approx0.5\)) and a status like \textit{investigating} or \textit{supported}. This conservative posture reduces false positives and keeps the trail of reasoning intact, but it also implies manual review for some items in the final deliverable. In practice, teams often triage these open leads using the same graph context and evidence slices, turning undecided claims into quick confirmations or rejections.

\section{Benchmarks}
We evaluate on ScaBench\cite{scabench} (real audit findings from Code4rena, Cantina, Sherlock) and score with Nethermind's AuditAgent matching algorithm\cite{auditagent2025}. For five shared projects \textit{(cantina\_minimal-delegation\_2025\_04, cantina\_smart-contract-audit-of-tn-contracts\_2025\_08, code4rena\_kinetiq\_2025\_07, code4rena\_lambowin\_2025\_02, code4rena\_secondswap\_2025\_02)}, Hound outperforms the baseline analyzer on recall and F1 at a modest precision trade-off.

\textbf{Shared-project micro results (Hound vs. baseline).} Truth=109 issues.
\begin{itemize}
  \item True Positives: 34 vs. 9 (\(+25\); FNs: 75 vs. 100)
  \item Precision: 9.3\% vs. 12.2\% (\(-2.9\) pp); Recall: 31.2\% vs. 8.3\% (\(+22.9\) pp; \(\approx 3.8\times\))
  \item F1: 14.2% vs. 9.8% (\(+4.4\) pp; \(+44\%\)); with partials: 16.3% vs. 11.9% (\(+4.4\) pp; \(+37\%\))
  \item Predictions per TP: 369/34 \(\approx 10.9\) vs. 76/9 \(\approx 8.4\)
\end{itemize}

\begin{center}
\scriptsize
\setlength{\tabcolsep}{4pt}
\resizebox{\linewidth}{!}{
\begin{tabular}{lrrrrrr}
\hline
Repo & Truth & Scan B/H & TP B/H & P\% B/H & R\% B/H & F1\% B/H \\
\hline
cantina\_minimal-delegation & 17 & 10 / 42 & 0 / 2 & 0.0 / 4.8 & 0.0 / 11.8 & 0.0 / 6.8 \\
cantina\_smart-contract-of-tn & 23 & 15 / 66 & 3 / 9 & 20.0 / 13.6 & 13.0 / 39.1 & 15.8 / 20.2 \\
code4rena\_kinetiq & 25 & 22 / 128 & 3 / 5 & 13.6 / 3.9 & 12.0 / 20.0 & 12.8 / 6.5 \\
code4rena\_lambowin & 14 & 14 / 54 & 2 / 5 & 14.3 / 9.3 & 14.3 / 35.7 & 14.3 / 14.7 \\
code4rena\_secondswap & 30 & 15 / 79 & 1 / 13 & 6.7 / 16.5 & 3.3 / 43.3 & 4.4 / 23.9 \\
\hline
\textbf{ALL (micro)} & 109 & 76 / 369 & 9 / 34 & 11.8 / 9.2 & 8.3 / 31.2 & 9.7 / 14.2 \\
\hline
\end{tabular}}
\end{center}

\textbf{Per-project recall uplift (TPs \(\to\) recall).}
\begin{itemize}
  \item cantina\_minimal-delegation: 0\(\to\)2 TPs; 0.0\%\(\to\)11.8\%
  \item cantina\_smart-contract-of-tn: 3\(\to\)9 TPs; 13.0\%\(\to\)39.1\%
  \item code4rena\_kinetiq: 3\(\to\)5 TPs; 12.0\%\(\to\)20.0\%
  \item code4rena\_lambowin: 2\(\to\)5 TPs; 14.3\%\(\to\)35.7\%
  \item code4rena\_secondswap: 1\(\to\)13 TPs; 3.3\%\(\to\)43.3\%
\end{itemize}

\textbf{Interpretation.} Flexible, analyst-defined graphs enable exact, cross-component retrieval, which substantially increases useful coverage (recall) across heterogeneous codebases. The belief system (long-lived hypotheses with confidence and explicit evidence) disciplines exploration and review. Together, these raise F1 (macro from 9.5\% to 14.4\%) with a modest precision trade-off typical of exploratory audits.

\textbf{Room for improvement.}
\begin{itemize}
  \item FP reduction without recall loss: (i) learned near-duplicate collapse over paraphrased hypotheses; (ii) stricter evidence gates (e.g., require two independent evidence slices or a cross-graph justification) before emission; (iii) lightweight severity calibration to drop weak low-impact leads.
  \item Early PoC/test integration: add coding-agent capabilities to author and execute sandboxed tests or minimal PoCs \emph{during} the audit stage (not only at finalization) so that incorrect hypotheses are rejected quickly and do not propagate downstream.
  \item Partial\(\to\)exact promotion: align titles/categories to canonical labels expected by the scorer to convert partials into matches (Hound: 5 partials across these projects).
  \item Efficiency: prioritize contradiction-rich components and reuse belief ledger to avoid re-probing low-signal areas, targeting \(<\!8\) predictions/TP while maintaining \(\ge\!30\%\) recall.
\end{itemize}

\section{Challenges and Possible Improvements}
Benchmarks show strong recall uplift and higher F1, but precision remains modest: the system still emits many plausible, ultimately incorrect hypotheses. We target earlier, stricter verification while preserving coverage:

\begin{itemize}[leftmargin=*]
  \item \textbf{Early verification gates:} Collapse the audit and finalization stages into a tighter single loop that attempts verification \emph{before} emission. Require cross-graph justification or two independent evidence slices; demote or drop items that fail checks instead of sending them to downstream QA.
  \item \textbf{Test-driven confirmation:} Enable the agent to write and run hermetic, sandboxed tests and minimal PoCs during analysis (e.g., unit/fuzz tests for contracts), attaching logs and artifacts as evidence. Passing tests raise confidence; failing tests refute hypotheses quickly and suppress repeats.
  \item \textbf{Learned deduplication and calibration:} Collapse near-duplicate hypotheses, normalize titles/categories to the scorer’s taxonomy, and calibrate severity to deprioritize low-impact leads without harming recall.
  \item \textbf{Negative-evidence memory:} Record strong refutations in the belief ledger and propagate them across sessions to avoid re-probing low-signal areas.
  \item \textbf{Efficiency targets:} Reduce predictions per TP from \(\approx 10.9\) to \(\le 8\) while sustaining \(\ge 30\%\) recall by prioritizing contradiction-rich components and reusing coverage state.
\end{itemize}

\paragraph{Open paper tasks.} Remaining work (see \texttt{paper/PAPER\_REFACTOR\_CHECKLIST.md}): (i) transcribe outline/deltas into the paper; (ii) references research pass (claims-to-cite list; citation improvements; bibliography standardization); (iii) compile/style sweep (overfull boxes, URL breaks).

\section*{Artifacts and Reproducibility}
We release a fully functional audit agent tool to support reproduction and further research \cite{hound}. All results can be reproduced using the SCABench curated dataset\cite{scabench} and the AuditAgent-based scorer\cite{auditagent2025}.

\end{document}